\documentclass[usenatbib]{mnras}

\newcommand{\be}{\begin{equation}}
\newcommand{\ba}{\begin{eqnarray}}
\newcommand{\ee}{\end{equation}}
\newcommand{\ea}{\end{eqnarray}}

\usepackage{graphicx}	
\title[]{Reconstruction of CMB Temperature Anisotropies with Primordial CMB Induced Polarization in Galaxy Clusters}

\author[G.-C. Liu et al.]{
Guo-Chin Liu,$^{1}$\thanks{E-mail: liugc@mail.tku.edu.tw(G.-C. Liu)}
Kiyotomo Ichiki,$^{2}$
Hiroyuki Tashiro$^{3}$
Naoshi Sugiyama$^{2,4,5}$
\\
$^1$Department of Physics, Tamkang University, Tamsui, New Taipei City 25137, Taiwan\\
$^2$Kobayashi-Maskawa Institute for the Origin of Particles and the Universe,
Nagoya University, Chikusa-ku, Nagoya, 464-8602, Japan\\
$^3$Department of Physics, Graduate School of Science, Nagoya University, Aichi 464-8602, Japan\\
$^4$Department of physics and astrophysics, Nagoya University, Nagoya 464-8602, Japan\\
$^5$Kavli Institute for the Physics and Mathematics of the Universe (WPI), Todai Institutes
for Advanced Study, The University of Tokyo, Kashiwa 277-8568, Japan}

\begin{document}
\label{firstpage}
\pagerange{\pageref{firstpage}--\pageref{lastpage}}
\maketitle
\begin{abstract}
Scattering of cosmic microwave background (CMB) radiation in galaxy clusters induces polarization signals determined by the quadrupole anisotropy in the photon distribution at the location of clusters. This "remote quadrupole" derived from the measurements of the induced polarization in galaxy clusters provides an opportunity of reconstruction of local CMB temperature anisotropies. In this {\em Letter} we develop an algorithm of the reconstruction through the estimation of the underlying primordial gravitational potential, which is the origin of the CMB temperature and polarization fluctuations and CMB induced polarization in galaxy clusters. We found a nice reconstruction for the quadrupole and octopole components of the CMB temperature anisotropies with the assistance of the CMB induced polarization signals. The reconstruction can be an important consistency test on the puzzles of CMB anomaly, especially for the low quadrupole and axis of evil problems reported in WMAP and Planck data.
\end{abstract}

\section{Introduction}
Large-scale anomalies have been reported in Cosmic Microwave Background (CMB) temperature map with several independent observations. A low quadrupole of CMB temperature anisotropies was first found in COBE data~\citep{Hinshaw} then confirmed by WMAP~\citep{Spergel}. The so-called axis of evil, which is an unusual alignment of the preferred axes of the quadrupole and octopole are found by several authors~\citep{axis_of_evil, Land, Samal}. Other anomalies include power asymmetry in north/south hemisphere~\citep{asymmetry, Hansen} and an anomalous cold spot~\citep{cold}. If the anomalies are not caused by foreground residuals or systematic effects, we are facing a challenge of understanding of fundamental physics and the nature of the cosmos. 

The polarization of the CMB is expected to provide valuable information on the nature of CMB anomalies. It is generated through Thomson scattering of temperature anisotropies on the last scattering surface~\citep{Hu}. If the anomalies in the CMB temperature are primordial, the polarization should thus exhibit similar peculiarities~\citep[see][for examples related to the Cold Spot, axis of evil, respectively]{Vielva, Frommert}.

One of the question here is that: Is it possible to reconstruct CMB temperature map for the low multipole $l$ from other independent observations?  If the answer is yes, then the reconstructed temperature map may be used to distinguish whether the anomalies are primordial or systematic. CMB polarization is one of the candidates. Due to the poor  correlation coefficients (less than 0.5 for multipole $l=2$ and $l=3$), however, the reconstruction from only CMB polarization can not achieve our goal. The scattering of CMB photons in clusters of galaxies may shed light on this attempt. 

The scattering of CMB photons in clusters of galaxies induces a polarization signal, which is determined by the quadrupole anisotropy in the photon distribution at the cluster location~\citep{Sazonov}. Therefore, the remote quadrupole in distant clusters, in principle, is investigable by the measurements of this induced polarization signal. The measurements of these signals were originally proposed to suppress the cosmic variance uncertainty~\citep{Kamionkowski, Bunn, Portsmouth}. Moreover, the magnitude of these signals gives some clues about the evolution of the CMB quadrupole to probe the dark energy. Of particular interest here is that it probes three-dimensional information of potential fluctuations around our last scattering surface~\citep{Seto}. Even though this induced polarization in galaxy clusters has not beed observed, its detection contains rich information for study of cosmology. 

In this {\em{Letter}} we propose an independent observation for the study of the CMB anomalies. We explore a practically useful cosmological probe from the measurements of remote quadrupole: the reconstruction of CMB temperature anisotropies for low multipole. Provided by the strong correlation of the remote quadrupole in low redshifts with the local CMB~\citep{Hall}, the reconstructed CMB temperature anisotropies are accurate enough for distinguishing the sources of low quadrupole and axis of evil problems. 

\section{Simulations of CMB Sky and CMB Induced Polarization in Distant Galaxy Clusters}

The CMB temperature anisotropies and E-mode polarization from a single plane wave can be written as~\citep{Ma}
\begin{equation}
\Delta_X({\bf{k}},\eta, {\bf{\hat{n}}})=\sum_l
(-i)^l(2l+1)\Delta_{Xl}({\bf{k}},\eta)P_l(\hat{\bf{k}}\cdot\hat{\bf{n}}), 
\label{eq:DeltaT}
\end{equation}
where $P_l$ is Legendre polynomial,  $\Delta_{Xl}$ is the transfer function with $X=T, E$ for temperature anisotropies or $E$-mode polarization, respectively. To compute the spherical harmonics coefficients of the temperature anisotropy and $E$-mode polarization $a_{X,lm}$ and some derivations later, we need the additional theorem of the spin-weighted spherical harmonics~\citep{Ng_Liu}
\ba
&&\sum_m \ _{s_1} Y_{lm}^*(\hat{\bf{n}}')\ _{s_2}Y_{lm}(\hat{\bf{n}}) \nonumber \\
&& =\sqrt{\frac{2l+1}{4\pi}}(-1)^{s_1-s_2} \ _{-s_1}Y_{ls_2}(\beta,\alpha)e^{-is_1\gamma},
\label{eq:addtheo} 
\ea
where $\alpha$, $\beta$ and $\gamma$ are the Euler angles as being composed of a rotation $\alpha$ around $\hat{\bf{e}}_3$, followed by $\beta$ around the new $\hat{\bf{e}}_2^{\prime}$ and finally $\gamma$ around $\hat{\bf{e}}_3^{\prime\prime}$. Then the spherical harmonics coefficients are
\be
a_{X,lm}=(-i)^l4\pi\int d^3k Y^*_{lm}(\hat{\bf{k}}) \Delta_{Xl}(\bf{k},\eta).
\label{eq:alm}
\ee

On the other hand, the polarization effect in the distant clusters of galaxies  arises from the presence of the quadrupole component of the CMB in the rest-frame of a cluster. For a cluster located in the $\hat{\bf{z}}$ direction, the primordial CMB quadrupole induced polarization is ~\citep{Ramos}
\begin{equation}
(Q_T\pm iU_T) = \sqrt{\frac{2\pi}{15}}\tau\int{d^3 {\bf k}  e^{i{\bf
      k}\cdot{\bf x} } \Delta_{T2}({\bf k}, \eta)} Y_{2\mp 2}(\hat{\bf{k}}),  
\label{eq:dqqicmb} 
\end{equation}
where $\tau$ is the optical depth across the cluster, $Q_T$ and $U_T$ are Stokes parameters in the unit of brightness temperature.  We adopt here, by convection, $Q_T<0$ ($Q_T>0$) for a N-S (E-W) polarization component and $U_T>0$ ($U_T<0$) for a NE-SW (NW-SE) component. We have assumed that free electrons in a cluster see the same CMB quadrupole because the primordial CMB temperature quadrupole has variations on much larger scales than the extent of individual clusters~\citep{Ramos}. Using the addition theorem of spin-weighted spherical harmonics in Eq.(\ref{eq:addtheo}), we rewrite Eq.~(\ref{eq:dqqicmb}) for a cluster located in any line of sight direction $\hat{\bf{n}}$ 
\begin{eqnarray}
(Q_T\pm iU_T) &=& \frac{2\sqrt{6}\pi}{5}\tau\int{d^3 {\bf k}  e^{i{\bf
      k}\cdot{\bf x} } \Delta_{T2}({\bf k}, \eta)}  \nonumber \\ 
      && \sum_m Y^*_{2m}(\hat{\bf{k}})\ _{\mp 2} Y_{2m}(\hat{\bf{n}}).
\label{eq:dqqicmb} 
\end{eqnarray}

All the CMB temperature anisotropy and its polarization and the polarization signal at distant clusters of galaxies can be computed directly from Eqs.(\ref{eq:alm}) and (\ref{eq:dqqicmb}), provided that the $\Delta_{Xl}({\bf k}, \eta)$ for each wave-mode is known. 
Since the evolution of $\Delta_{Xl}({\bf k}, \eta)$ is independent 
of the direction of $\bf {k}$, we may write $\Delta_{Xl}({\bf k}, \eta)=\Delta_{Xl}( k, \eta) \Psi({\bf k})$,
where $\Psi({\bf k})$ is the primordial gravitational potential. This primordial gravitational potential $\Psi({\bf k})$ that appears in Eq.~(\ref{eq:alm}) and (\ref{eq:dqqicmb}) is the origin of CMB temperature and polarization fluctuations and primordial CMB induced polarization in galaxy clusters. Therefore, estimating the primordial gravitational potential in three-dimensional Fourier space is our first step for the reconstruction. 

It is usual to assume that the two-point correlation function of the $\Psi({\bf k})$  has the form 
\begin{equation}
<\Psi^*({\bf k})\Psi({\bf k^\prime})>=P_{\Psi}(k)\delta^3({\bf
  k}-{\bf k^\prime}), 
\label{eq:sigma2}  
\end{equation}
and the power spectrum $P_{\Psi}(k)$ obeys a power-law
\begin{equation}
P_{\Psi}(k)=Ak^{n_s-4}
\label{eq:power_law}
\end{equation}
with $A$ being a normalization factor and $n_s$ a spectral index of the scalar
perturbations.


To compute the simulation data, we first generate the three-dimensional primordial gravitational potential $\Psi({\bf k})$ by drawing a random number from a gaussian distribution with variance $P_{\Psi}(k)$ for each wave-mode. The time evolution of the transfer function $\Delta_{Xl}(k, \eta)$ is calculated numerically by the CMBFast~\citep{CMBFast} Boltzmann code, assuming the cosmological parameters from the Planck 2013 results~\citep{Planck_2014}. Different skies can be generated by changing the seed of our random number generator routine. 
Given the simulated primordial gravitational potential, we generate CMB data $a_{T,lm}$ and $a_{E,lm}$ using Eq.(\ref{eq:alm}) and polarization data in clusters $Q_T$ and $U_T$ using Eq.~(\ref{eq:dqqicmb}). The observed CMB induced polarization in clusters is linear proportional to the cluster's optical depth, which may be extracted from X-ray surface brightness observations if the temperature profile is known. Here we simply assume $\tau=1$. The position of the galaxy clusters is assumed to be randomly distributed in the universe from $z=0$ to $4$.  We perform the integration in Eqs.(\ref{eq:alm}) and (\ref{eq:dqqicmb}) by summing the contribution from each Fourier mode in spherical coordinates. In each radial direction of ${\bf k}$, we have sampled uniformly 240 modes in logarithm space from $k=7 \times 10^{-6}$ to $1.4 \times 10^{-1}$ $h$/Mpc. The angular directions of $(\theta_k, \phi_k)$ are obtained by the Healpix scheme~\citep{healpix} at resolution-5 map.

\begin{figure}
\centering
\includegraphics[width=5cm, angle=90]{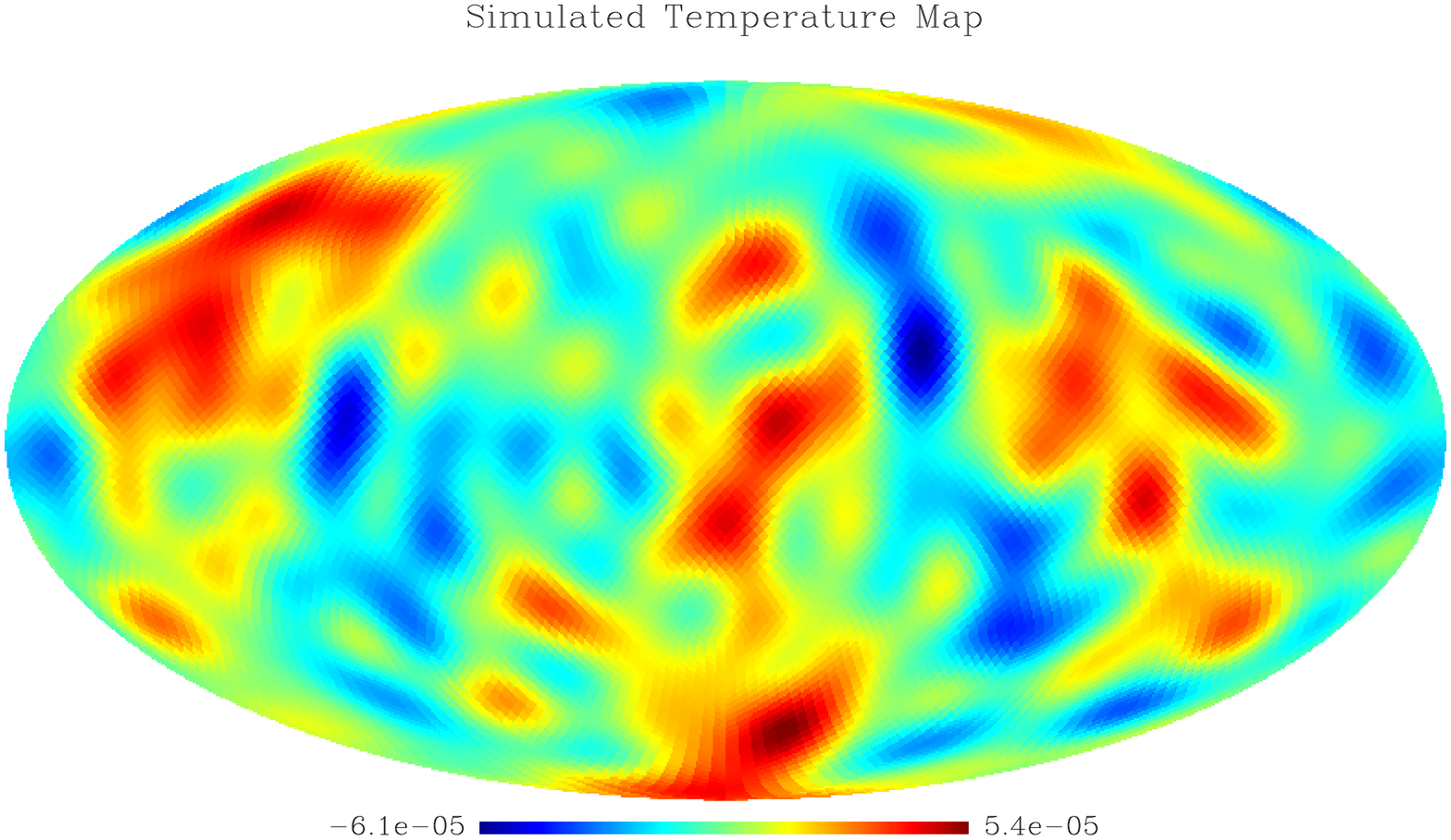}
\includegraphics[width=5cm, angle=90]{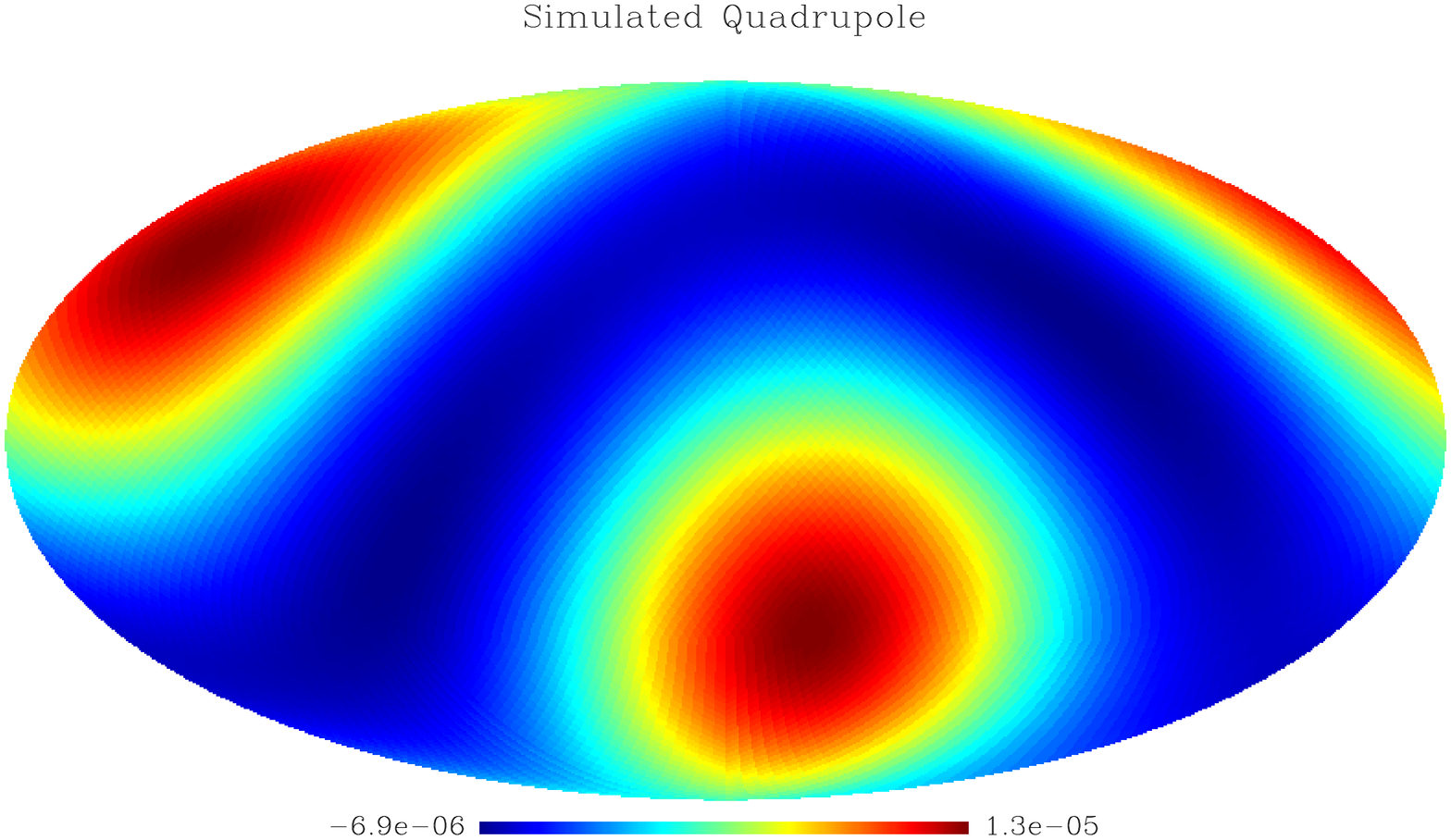}
\includegraphics[width=5cm, angle=90]{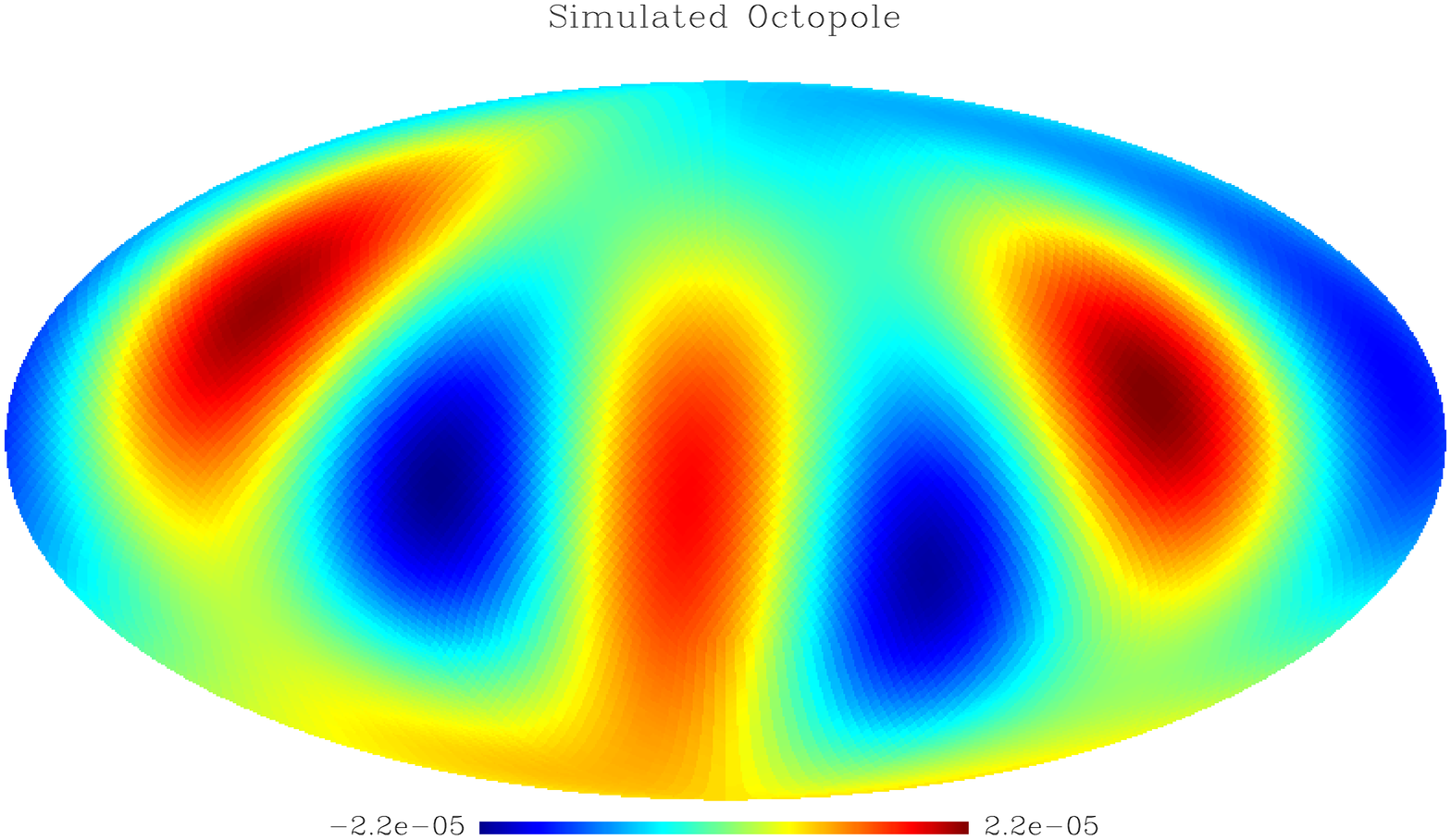}
\caption{Simulated CMB temperature map (upper) and its quadrupole (middle) and octopole (lower) components.} \label{cmbmaps}
\end{figure}

\section{Reconstruction of CMB Temperature Anisotropies}

We present in Fig.~(\ref{cmbmaps}) a typical sky realization of the CMB temperature anisotropies ($l\le16$) and its quadrupole and octopole components. In order to reconstruct this CMB temperature map with CMB induced polarization in galaxy clusters from the same realization, we first {\em estimate} the primordial gravitational potential using Bayes' theorem
\begin{equation}
Pr(\Psi|{\bf d})\propto Pr({\bf d}|\Psi) Pr(\Psi),
\end{equation}
where $Pr(\Psi)$  is the prior probability, $Pr({\bf d}|\Psi)$ 
is the likelihood of obtaining the data ${\bf d}=\{Q_T, U_T\}$ given a realization of primordial gravitational potential $\Psi$.  The estimation of $\Psi$ is obtained by minimizing the function
\begin{eqnarray}
f&=&\sum_{j=1}^{n_c}\frac{(\hat{Q}_{T,j}-Q_{T,j})^2}{2\sigma_{Q_T}^2} +
\sum_{j=1}^{n_c}\frac{(\hat{U}_{T,j}-U_{T,j})^2}{2\sigma_{U_T}^2}\nonumber\\ 
& & + \sum_{k=1}^{n_k}\frac{\Psi_k^2}{2P_{\Psi}},
\label{eq:mini}
\end{eqnarray}
where $\sigma_{Q_T}$ and $\sigma_{U_T}$ are instrument noise of observation, $n_k$ is the number of Fourier modes in the fitting, $n_c$ is the number of observed clusters of galaxies, $\Psi_k$ is the discretized $\Psi(k)$, $\hat{Q}_{T,j}$ and $\hat{U}_{T,j}$ are the reconstructed Stokes parameters in the $j$th galaxy cluster with the estimated $\Psi$. The last term in the right side is the logarithm of the prior. 

\begin{figure}
\centering
\includegraphics[width=5cm, angle=90]{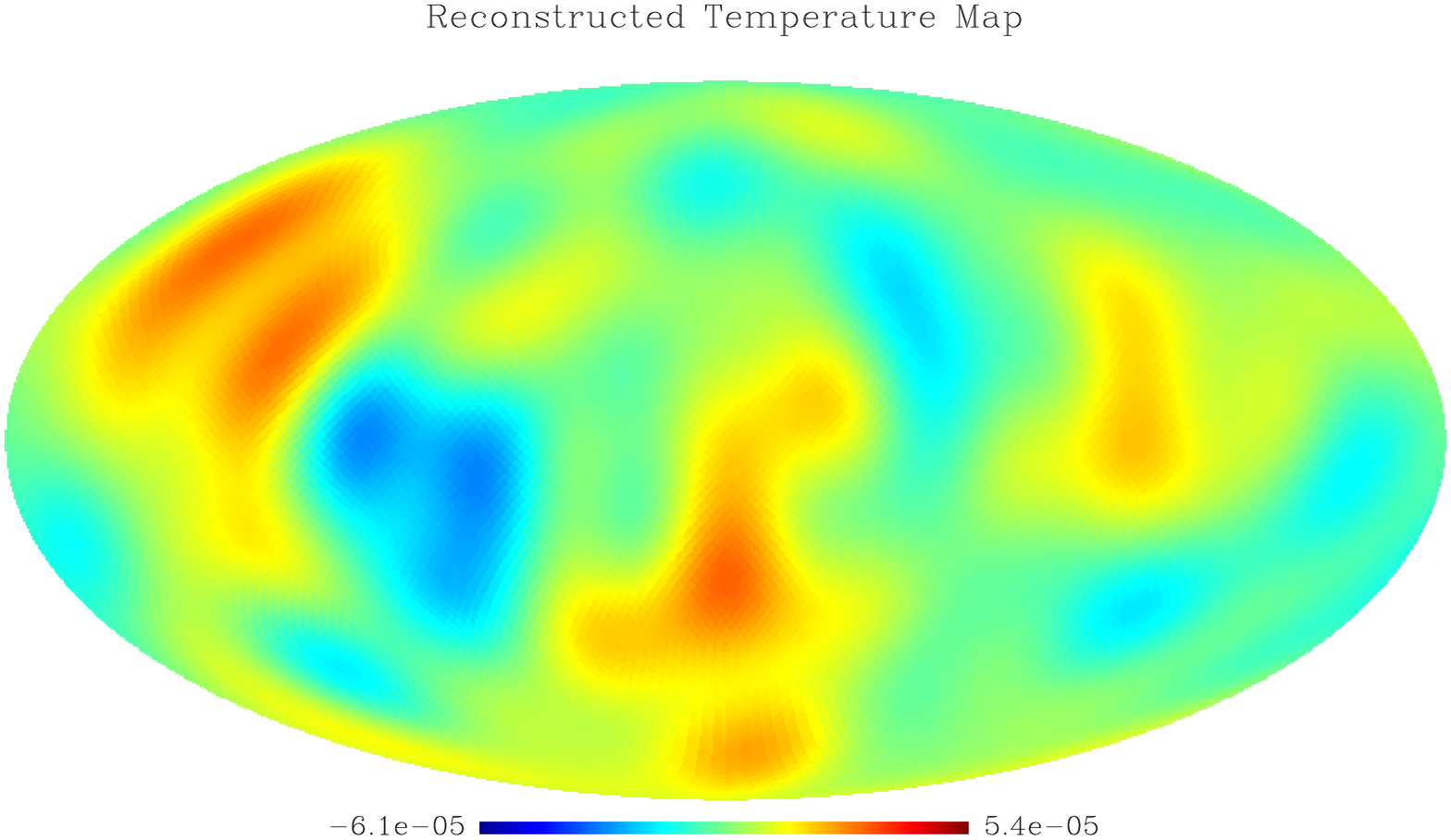}
\includegraphics[width=5cm, angle=90]{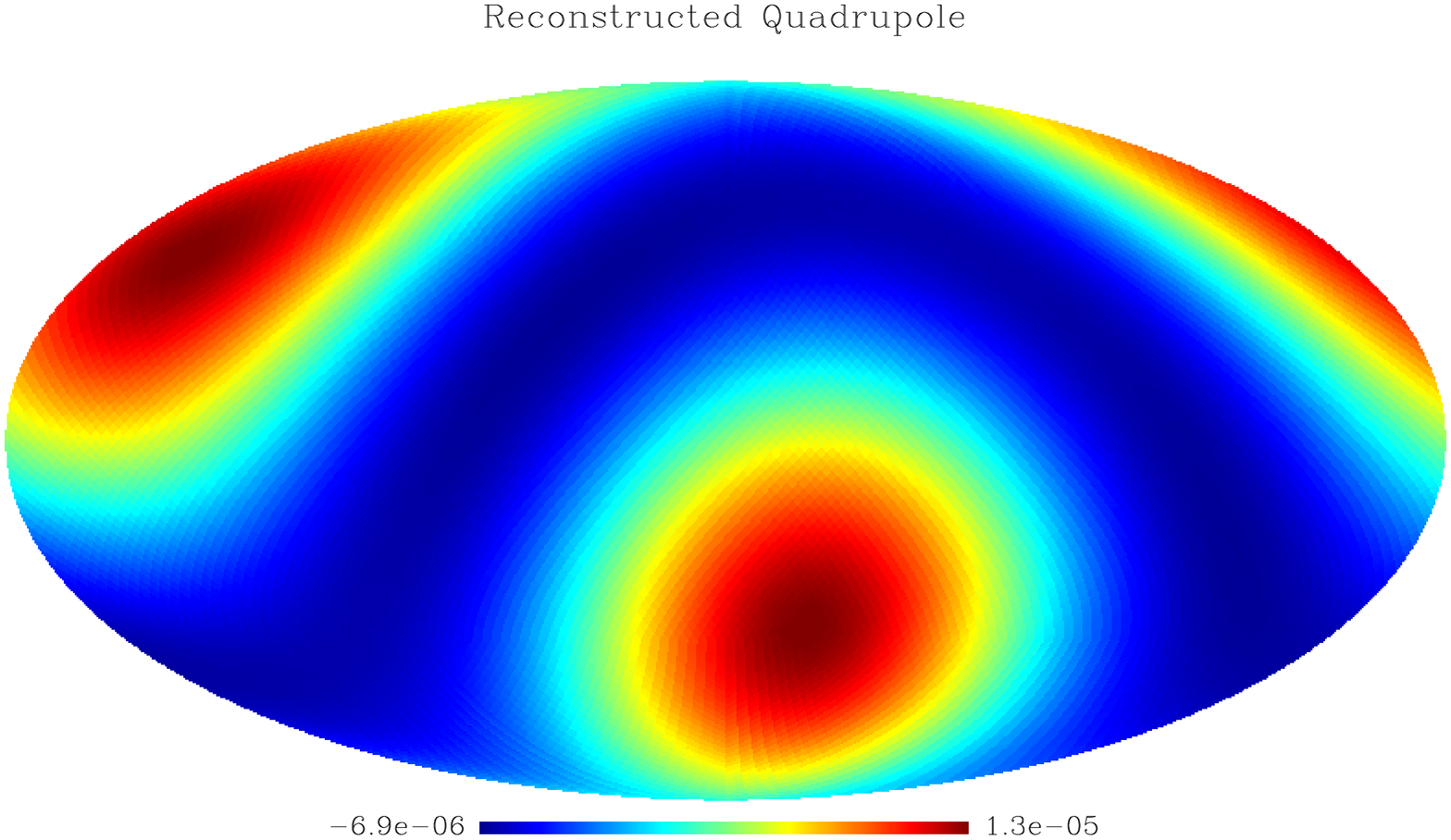}
\includegraphics[width=5cm, angle=90]{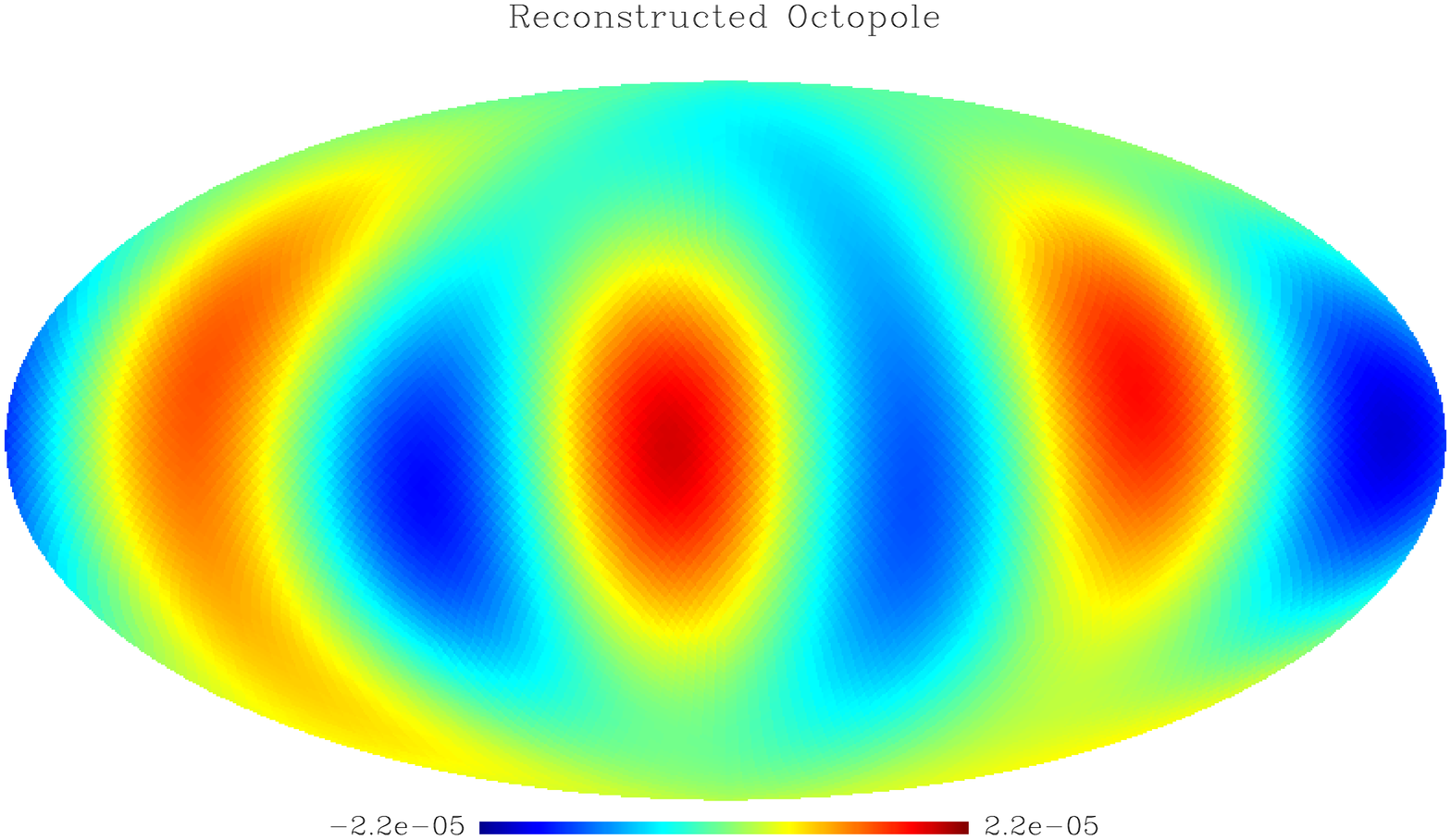}
\caption{Reconstructed CMB temperature anisotropies (upper) and its quadrupole (middle) and octopole (lower) components.} \label{fig:recmaps}
\end{figure}

There must be many possible sets of $\Psi$ corresponding to one observation within the desired uncertainty since the number of modes $n_k$ is much bigger than the number of data $n_c$. Therefore, the reconstructed $\Psi$ differ with the input, while the reconstructed Stokes parameters in the galaxy clusters do not change too much.  However, reducing $n_k$ to avoid the overfitting produces another problem. It generates the spurious correlation between the induced polarization in clusters and the CMB temperature anisotropy and $E$-mode polarization once we reduce the number of modes $n_k$ in radial direction. Moreover, the orthogonality property of the spherical harmonics may be broken once we reduce the number of modes $n_k$ in angular direction. 
When we perform the minimization, we use a smaller number of $n_k$ than the input. We reduce the number of sampling from 240 to 60 for the radial direction, and number of pixel from Healpix resolution-5 to resolution-3 map for the angular direction of ${\bf k}$. The total number of the parameters in the fitting is $46080$ (complex variables). The relative errors of $Q_T$, $U_T$ and $a_{X,lm}$ with $l \le 10$ in this configuration are less than $3 \%$. We also tested bigger $n_k$ but it does not significantly change our conclusion below. 

\begin{figure}
\centering
\includegraphics[width=8cm]{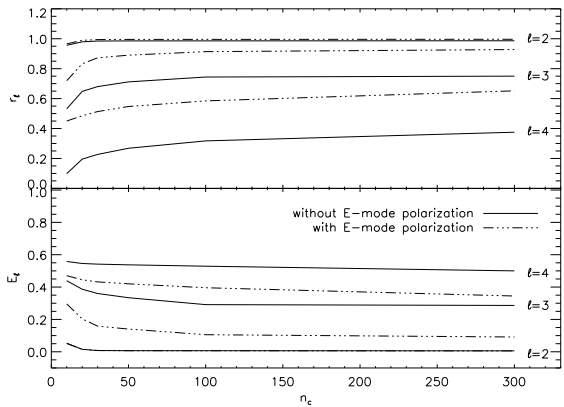}
\caption{Forecast of $r_l$ (upper) and $E_l$ (lower) for $l=2,3,4$  as function of number of observing galaxy clusters without (solid) and with (dashed) additional $E$-mode polarization data. The redshift observed galaxy clusters uniformly distributes from $z=0$ to $z=1.0$.} \label{fig:corr_err}
\end{figure}

\section{Conclusions}

In Fig.~(\ref{fig:recmaps}) we present the reconstructed temperature map ($l \le 10$) and its quadrupole and octopole components with the estimated $\Psi$, which is obtained by fitting to the simulated data from 300 galaxy clusters whose redshifts $z<1.0$ in our catalog. It is obvious that the reconstructed temperature map loses power on small angular scales. However, with the assistance of strong correlation of remote quadrupole with local CMB, the quadrupole component is almost completely reconstructed and the octopole component is also similar to the simulated one. It shows the feasibility of the reconstruction with our algorithm and the results should be useful for the study of anomalies in CMB maps. For the higher multipole, the reconstruction becomes worse due to the more deviation of the transfer functions from that of remote quadrupole with increasing $l$.

\begin{figure}
\centering
\includegraphics[width=8cm]{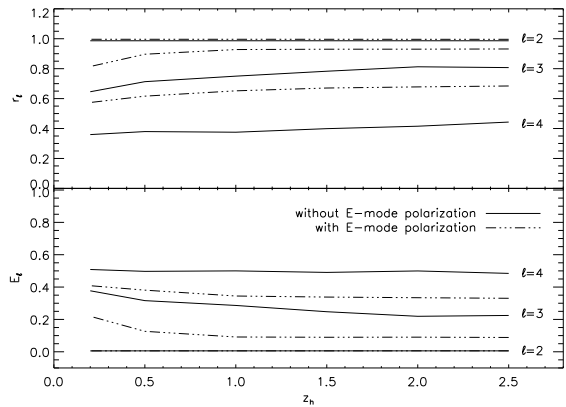}
\caption{Forecast of $r_l$ (upper) and $E_l$ (lower) for $l=2,3,4$ for the depth of the observation without (solid) and with (dashed) additional $E$-mode polarization data. The number of observed galaxy clusters $n_c$ is fixed to 300.} \label{fig:effect_z}
\end{figure}

We forecast the observation strategy for the induced polarization in galaxy clusters. To quantify the  goodness of the reconstruction, we defined the dimensionless correlation coefficient $r_l=\sum_m a_{T,lm}\hat{a}_{T,lm}/(2l+1)\sqrt{C_l\hat{C}_l}$ and error of reconstruction $E_l=\sum_m (\hat{a}_{T,lm}-a_{T,lm})^2/(2l+1)C_l$ with $C_l$ the input temperature power spectrum from simulation. We make 100 realizations and select the observed galaxy clusters with redshifts $z \le z_h$. We show the average of $r_l$ and $E_l$ as function of $n_c$ for $l=2, 3$ and $4$ in Fig~(\ref{fig:corr_err}) with fixed $z_h=1$. With tens of observed galaxy clusters the quadrupole component can be perfectly reconstructed and be useful for the investigation of low quadrupole problem. The reconstruction of higher multipole components improves with the increasing number of observed galaxy clusters. We also test the reconstruction by adding $E$-mode polarization data (dashed curves).  With the additional data $a_{E,lm}$ for $l \le 10$, the reconstruction of octopole and higher multipole anisotropies improves significantly through the $TE$ correlation. The relation of the reconstruction and the depth of the observation is shown in Fig.~(\ref{fig:effect_z}). The remote quadrupole of observed in high redshift galaxy clusters probes the universe on scales smaller to the local CMB, so the reconstruction of octopole and higher multipole improves with increasing $z_h$.

The procedure we have here is something of an idealization. We assume the optical depth through the galaxy clusters is precisely measured and the instrumental noise is significantly small ($\sigma_{Q_T}=\sigma_{U_T}=\sigma_{lm}=10^{-3} \mu K$). We also ignore all the contaminations of the polarization signal, for example, the polarization induced by the transverse peculiar velocity of the galaxy clusters~\citep{Sazonov, Ramos} and the background polarization. The signal is probably detectable in the next-generation polarization experiments with broad frequency coverage such as PRISM~\citep{prism} but the separation of the quadrupole signal from the other contaminants would be an experimental challenge. 

\section*{Acknowledgements}

This work has been supported in part by the Ministry of Science and Technology, Taiwan, ROC under the Grants No. 104-2112-M-032 -007 - (G.-C. L.) and Grant-in-Aid for Scientific Research from the Ministry of Education, Culture, Sports, Science and Technology (MEXT), Japan under the Grants Nos.
15H05890 (N.S. and K.I.),  24340048 (K. I.) and 15K17646 (H.T.).
H.T. also acknowledges the support by MEXT's Program for Leading Graduate Schools PhD professional, "Gateway to Success in Frontier Asia".


\begin{thebibliography}{99}
\bibitem[\protect\citeauthoryear{Bunn}{2006}]{Bunn} Bunn E. F., \prd, 73, 123517
\bibitem[\protect\citeauthoryear{Cruaz et al.}{2006}]{cold} Cruz M. et al., 2006, \mnras, 369, 57
\bibitem[\protect\citeauthoryear{de Oliveira-Costa et al.}{2004}]{axis_of_evil} de Oliveira-Costa A. et al., 2004, Phys. Rev. D, 69, 063516
\bibitem[\protect\citeauthoryear{Eriksen et al.}{2004}]{asymmetry} Eriksen H. K. et al., 2004, \apj, 605, 14
\bibitem[\protect\citeauthoryear{Frommert and Ensslin}{2010}]{Frommert} Frommert M. and Ensslin T. A., 2010 \mnras, 403, 1739
\bibitem[\protect\citeauthoryear{Gorski et al.}{2005}]{healpix} Gorski K. M. et al., 2005,  \apj,  622, 759
\bibitem[\protect\citeauthoryear{Hall and Challinor}{2014}]{Hall} Hall A. and Challinor A., 2014, \prd, 90, 063518
\bibitem[\protect\citeauthoryear{Hansen et al.}{2009}]{Hansen} Hansen F. K. et al., 2009, \apj, 704, 1448
\bibitem[\protect\citeauthoryear{Hinshaw et al.}{1996}]{Hinshaw} Hinshaw G. et al., 1996, Astrophysical Journal Letters, 464, L25 
\bibitem[\protect\citeauthoryear{Hu and White}{1997}]{Hu} Hu W. and White M., 1997, \na, 2, 323
\bibitem[\protect\citeauthoryear{Kamionkowski and Loeb}{1997}]{Kamionkowski} Kamionkowski M. and Loeb A., 1997, Phys. Rev. D, 56, 4511
\bibitem[\protect\citeauthoryear{Land and Magueijo}{2005}]{Land} Land K. and Magueijo J., 2005, Phys. Rev. Lett., 95, 071301
\bibitem[\protect\citeauthoryear{Ma and Bertschinger}{1995}]{Ma} Ma C. and Bertschinger E., 1995, \apj, 455, 7
\bibitem[\protect\citeauthoryear{Ng and Liu}{1999}]{Ng_Liu}  Ng K.-W. and Liu G.-C., 1999, Int. J. Mod. Phys. D8, 61
\bibitem[\protect\citeauthoryear{Planck Collaboration et al.}{2014}]{Planck_2014} Planck Collaboration et. al., 2014, A\& A, 571, 16
\bibitem[\protect\citeauthoryear{Portsmouth}{2004}]{Portsmouth} Portsmouth J., 2004, Phys. Rev. D, 70, 063504
\bibitem[\protect\citeauthoryear{Prism Collaboration et al.}{2013}]{prism} PRISM Collaboration et al., ArXiv e-prints, 1306.2259 (2013)
\bibitem[\protect\citeauthoryear{Ramos, da Silva and Liu}{2012}]{Ramos} Ramos E. P. R. G., Silva A. da and Liu  G.-C., 2012, \apj, 757, 44
\bibitem[\protect\citeauthoryear{Samal et al.}{2008}]{Samal} Samal P. K. et al., 2008, \mnras, 385,1718
\bibitem[\protect\citeauthoryear{Sazonov and Sunyaev}{1999}]{Sazonov} Sazonov S. Y. and Sunyaev R. A., 1999, Mon. Not. R. Astron. Soc. 310, 765
\bibitem[\protect\citeauthoryear{Seljak and Zaldarriaga}{1996}]{CMBFast} Seljak U. and Zaldarriaga M., 1996, \apj, 469, 437
\bibitem[\protect\citeauthoryear{Seto and Sasaki}{2000}]{Seto} Seto N. and Sasaki M., 2000, Phys. Rev. D, 62, 123004
\bibitem[\protect\citeauthoryear{Vielva et al.}{2011}]{Vielva} Vielva P. et al., 2011, \mnras, 410, 33
\bibitem[\protect\citeauthoryear{WMAP Collaboration et al.}{2003}]{Spergel} WMAP Collaboration et al., 2003, \apjs, 148, 175
\end{thebibliography}
\end{document}